\documentclass[reprint,amsmath,amssymb,aps,longbibliography]{revtex4-1}

\usepackage{graphicx}
\usepackage{color}
\usepackage{amsmath}
\usepackage{textcomp}
\usepackage{epstopdf}
\usepackage{revsymb}
\usepackage{amsbsy}
\usepackage{mathrsfs}
\usepackage{amsfonts}
\usepackage{amsthm}
\usepackage{float}
\usepackage{natbib}
\usepackage{xcolor}
\usepackage[colorlinks=true, pdfborder={0 0 0}, linkcolor=red,citecolor=blue, anchorcolor=blue]{hyperref}
\newcommand{\ket}[1]{| #1 \rangle}

\theoremstyle{plain}

\theoremstyle{definition}

\begin{document}

\title{ Three Party Quantum Networks Created by Quantum Cloning}
\author{Manish Kumar Shukla}
\affiliation{Independent Researcher, Eden Au Lac Apartment, Indiranagar, Bangalore 560038, India}
\email{manish.shukla393@gmail.com}
\author{Minyi Huang}
\affiliation{Department of Mathematical Sciences, Zhejiang Sci-Tech University, Hangzhou 310018, PR China}
\email{11335001@zju.edu.cn}
\author{Indranil Chakrabarty}
\affiliation{Center for Security, Theory and Algorithmic Research, International Institute of Information Technology-Hyderabad, Gachibowli, Telangana-500032, India.}
\email{indranil.chakrabarty@iiit.ac.in}
 \author{Junde Wu}
 \affiliation{School of Mathematical Sciences, Zhejiang University, Hangzhou 310027, PR-China}
 \email{wjd@zju.edu.cn}

\begin{abstract}
With progress in quantum technologies, the field of quantum networks has emerged as an important area of research. In the last few years, there has been substantial progress in understanding the correlations present in quantum networks. %Although these studies are restricted to simple network topologies, nevertheless they present an important picture of genuine quantum correlations present exclusively in quantum networks. 
In this article, we study cloning as a prospective method to generate three party quantum networks which can be further used to create larger networks.  We analyze various quantum network topologies that can be created using cloning transformations. This would be useful in the situations wherever the availability of entangled pairs is limited. In addition to that we focus on the problem of distinguishing networks created by cloning from those which are created by distributing independently generated entangled pairs. We find that there are several states which cannot be distinguished using the Finner inequalities in the standard way. For such states, we propose an extension to the existing Finner inequality for triangle networks by further increasing the number of observers from three to four or six depending on the network topology. This takes into account the additional correlations that exist in the case of cloned networks. In the last part of the article we have used tripartite mutual information to distinguish cloned networks from networks created by independent sources and have further used squashed entanglement as a measure to quantify the amount of dependence in the cloned networks.

\end{abstract}
     
\maketitle

\section{Introduction}
Quantum correlations have never failed to amaze researchers and have remained an important aspect of quantum information theory \cite{einstein1935einstein,  heisenberg1927w, nielsen2001quantum, bell1964einstein}. In recent years, several studies were made to further enhance our understanding the nature of quantum correlations and other quantum resources. In particular a resource theoretic framework was constructed  to understand the role of these resources \cite{horodecki2009quantum, henderson2001classical, ollivier2001quantum, bera2017quantum, baumgratz2014quantifying} in different information processing tasks.
Also, efforts have been made to extend the idea of quantum correlations and other quantum resources to three or more parties \cite{greenberger1989going, chakrabarty2011quantum}. The idea is not only about quantum correlations in a multi-party scenario, but the reach also extends to quantum networks and internet \cite{wehner2018quantum,biamonte2019complex,chiribella2009theoretical,perseguers2010quantum,kozlowski2019towards,perseguers2013distribution,kimble2008quantum, simon2017towards,renou2019genuine}. These networks are in principle created by the distribution of entanglement to distant parties by one or multiple independent sources and application of local unitaries to establish strong correlations in the entire network. In such scenarios, distant parties can establish direct correlation through intermediate nodes by processes like repeater technology \cite{sazim2013study,li2020effective,zukowski1993event}. Whenever we talk about multi-party entanglement, monogamy plays the central role. In case of quantum networks, since they can be created by distributing independently created entangled pairs, monogamy is not that much applicable. Bell inequalities for such networks have also been studied in detail \cite{branciard2010characterizing, branciard2012bilocal, chaves2012entropic, tavakoli2014nonlocal} and even enhanced in the way that they can be input independent \cite{branciard2012bilocal, fritz2012beyond}. These ideas have further pushed our understanding towards real world quantum networks\cite{kimble2008quantum, simon2017towards}. However, the study of correlations in quantum networks is still not complete and in recent years we have seen tremendous theoretical and experimental progress in terms of establishment of quantum networks \cite{renou2019genuine}. Most of these studies are restricted to simple network topologies like bi-local and triangle networks, which are created using an independent source that prepares and distributes entangled qubits to distant observers\cite{perseguers2010quantum}.

\indent Quantum resources\cite{einstein1935einstein, ollivier2001quantum,henderson2001classical, baumgratz2014quantifying} on one hand, give us a significant advantage in accomplishing various information processing tasks \cite{bennett1984quantum, shor2000simple, ekert1991quantum, hillery1999m,sazim2015retrieving,adhikari2010probabilistic,ray2016sequential, bennett1993teleporting,horodecki1996teleportation, bennett1992communication, nepal2013maximally, horodecki2009quantum, bose1998multiparticle,zukowski1993event, sazim2013study, chatterjee2016broadcasting, sharma2017broadcasting, jain2019asymmetric, mundra2019broadcasting,  deutsch1996quantum} over their classical counterparts, while on the other hand, impose strict restrictions on certain kind of operations  \cite{wootters1982single, pati2000impossibility, patel2018impossibility, chakrabarty2007impossibility, pati2006no, modi2018masking}.
Among these restrictions, the No-cloning theorem prohibits us from perfectly cloning an arbitrary unknown quantum state \cite{wootters1982single}.
However, it does not rule out the possibility to clone states approximately (i.e. with fidelity of less than unity) \cite{wootters1982single, buvzek1996quantum, gisin1997optimal, buvzek1998universal, bruss1998optimal, cerf2000pauli, cerf2000nj, scarani2005quantum} or with a certain probability of success. Approximate quantum cloning machines (QCMs) can further be classified into two types, namely (a) state-dependent \cite{wootters1982single, buvzek1996quantum, bruss1998optimal, adhikari2006broadcasting, shukla2019broadcasting} and (b) state independent \cite{buvzek1996quantum, buvzek1998universal} cloners. Another kind of classification that can be made in context of these cloning machines is based on the kind of output they produce. If all the cloning outputs are same then the cloning machine is called symmetric, otherwise, it is referred to as an asymmetric cloning machine \cite{cerf2002security, ghiu2003asymmetric}. %Depending upon the task at hand one can select the most suited cloning machine from this arsenal.

%\textcolor{blue}{In this work, we assume that there is an independent source  which is trying to create a quantum network has a finite number of maximally entangled states. 
To start with, we make another assumption that the source has a single maximally entangled state and wants to create a tripartite quantum network. We also study the process of establishing a network without the need for an independent source.
We have considered the set of symmetric cloners, which produce outputs with equal fidelity \cite{bruss1998optimal}. We use such cloners to create various types of quantum networks and made an effort to  understand the limits of networks that are created by the process of cloning. One significant difference between networks created by cloning and by distributing independently created entangled states is the condition of statistical independence of resources, which relaxes the monogamy constraint and also makes the relevant bell inequalities to become non-linear in the latter case \cite{renou2019genuine}. We show that by choosing a specific initial states and cloning machine parameters, we can create networks which cannot be distinguished from those networks which are created by distributing independent entangled pairs. We thoroughly analyze such tripartite networks along with  various topologies that are created and study how these networks can be distinguished from the ones which are created by distributing independent entangled pairs. We find several instances where these cannot be distinguished. We also try to quantify the amount of dependence on the initial source in networks with the help of an entanglement quantifier. 

In section (II) we first introduce and discuss various types of cloning techniques that have been used to create different three party quantum networks. These include local and non-local cloning operations, with the option of creating either two copies or three copies. In Section III we briefly describe all types of tripartite network and also Finner inequality to understand the limit it poses on network topology.  We also introduce our extension to Finner inequality to understand these limits more. In section IV and try to create tripartite network topologies with cloning by taking inspiration from the work \cite{renou2019limits}. Specifically, we create three kinds of network topologies for three distant observers, namely tripartite networks, bi-local, and genuine triangle networks. We study the problem of distinguishing these networks from traditional networks. We also study the constraints on such networks arising due to the quantum Finner inequalities \cite{renou2019limits}. In section V we use a network witness created using tripartite mutual information to distinguish cloned networks from others since cloned networks are created using a single source. We also propose a method to quantify the statistical dependence on the initial single source present in the entire network. 

%We categorise networks into various types by studying the NPA conditions proposed in ref\cite{pozas2019bounding}.%

\section{Approximate Quantum Cloning Machines}
From the breakthrough paper from Wooters and Zurek we know that perfect cloning is not possible according to the No-cloning theorem \cite{wootters1982single}. However, it never rules out the possibility to clone a quantum state approximately with a fidelity $F$ less than unity\cite{buvzek1996quantum}. The Fidelity $F$ can be expressed as,
\begin{equation}
F = \left\langle \Psi|\rho^{out}|\Psi\right\rangle, 
\end{equation}
where $\left|\Psi\right\rangle$ refers to the state to be cloned at the input port of the cloner and $\rho^{out}$ is the state obtained at its output port after applying the cloning transformation. In our study of networks, we use similar symmetric approximate cloning machines. The base version of all the cloning machines used in this work is the Buzek-Hillery cloning machine.

Here, we have used four variants of Buzek–Hillery cloning machine to create three kinds of networks. The first cloning transformation is  local $1->2$ cloning($QCL_{1->2}$), where we apply two unitary transformations at the two ends of a spatially separated entangled pair, to produce another copy of the local states. It is given by the following transformation,
\cite{buvzek1998universal},
\begin{eqnarray}
&& U_{bh}\left|\Psi_i\right\rangle_{a_0} \left|\Sigma\right\rangle_{a_1} \left|X\right\rangle_x \rightarrow  c\left|\Psi_i\right\rangle_{a} \left|\Psi_i\right\rangle_{b} \left|X_{ii}\right\rangle_x \nonumber\\
&& +d\displaystyle \sum_{j\neq i}^{M} \left(\left|\Psi_i\right\rangle_{a} \left|\Psi_j\right\rangle_{b} +\left|\Psi_j\right\rangle_{a}\left|\Psi_i\right\rangle_{b}\right) \left|Y_{ij}\right\rangle_x.
\label{eq:B-H_gen_transform}
\end{eqnarray}
It is an $M$-dimensional quantum copying transformation acting on a state $\left|\Psi_i\right\rangle_{a_0}$ (where $i$ $\in$ \{1, ..., $M$\}). This state is to be copied on a blank state $\left|\Sigma\right\rangle_{a_1}$. Initially, the cloning machine was prepared in state $\left|X\right\rangle_x$, which after being applied by cloning transformation gets transformed into another set of state vectors $\left|X_{ii}\right\rangle_x$ and $\left|Y_{ij}\right\rangle_x$ (where $i,\,j$ $\in$ \{1, ..., $M$\}). Here, $\ket{\Psi} = \sum_{i=1}^{M}\alpha_i\ket{\Psi_i}$, where $\ket{\Psi_i}$ are the basis vectors of the m qubit system with dimensions $M = 2^m$ and $\alpha_i$ represents the probability amplitude, hence $\sum_{i=0}^{M} \alpha^{2}_i = 1$. The modes $a_0$, $a_1$ and $x$ represent the input, blank and machine qubits, respectively. In this case, these transformed machine state vectors $\left(\left|X_{ii}\right\rangle,\,  \left|Y_{ij}\right\rangle\right)$ are elements of the orthonormal basis set in the $M$-dimensional space. Here, $i,\:j$ are two indices that run from $1$ to $M$. The coefficients $c$ and $d$ are the probability amplitudes which take real values. The relation between $c$ and $d$ can be easily be obtained from the unitarity condition of cloning transformation, this relation is given by, $c^{2} = 1 - 2(m-1)d^2$. When we want to make use of state independent versions of the cloner, we can calculate the specific value of machine parameter $d$, for which the coning outcome is not dependent on input states.

The second cloning transformation is $1->3$ ($QCL_{1->3}$) cloning on a single qubit. Here we create three copies of a single qubit and distribute it to three parties, to create a three party network. This $1 \rightarrow 3$ cloning transformation applied on a single qubit is given by,

\begin{equation}
\begin{split}
    U_{bh}\ket{\Psi_{i}}_{a_0}\ket{\Sigma}_{a_1}
    \ket{\Sigma}_{a_2}\ket{X}_{x} 
 \rightarrow  
c \ket{\Psi_{i}}_{a} \ket{\Psi_{i}}_{b}\ket{\Psi_{i}}_{c} \ket{X_{ii}}_{x} \\
+ d \sum_{j,k\neq i}^{M} 
(\ket{\Psi_{i}}_{a}\ket{\Psi_{j}}_{b}\ket{\Psi_{k}}_{c} 
+ \ket{\Psi_{i}}_{a}\ket{\Psi_{k}}_{b}\ket{\Psi_{j}}_{c} \\
+\ket{\Psi_{j}}_{a}\ket{\Psi_{i}}_{b}\ket{\Psi_{k}}_{c}
+\ket{\Psi_{j}}_{a}\ket{\Psi_{k}}_{b}\ket{\Psi_{i}}_{c}\\
+\ket{\Psi_{k}}_{a}\ket{\Psi_{i}}_{b}\ket{\Psi_{j}}_{c}
+\ket{\Psi_{k}}_{a}\ket{\Psi_{j}}_{b}\ket{\Psi_{i}}_{c})
\ket{Y_{ijk}}_{x}.
\label{eq:B-H_gen_transform13}
\end{split}
\end{equation}

The third transformation is non local $1->2$ cloning transformation($QCNL_{1-2}$), applied together on an entangled pair to create another copy of it. We follow this mechanism of cloning when we want to create bi-local networks\cite{renou2019limits}, which is explained in the following sections. The cloning transformation is same as given in equation (\ref{eq:B-H_gen_transform}) where $M=4$. To create a triangle network we use the $1->3$ non-local cloning transformation ($QCNL_{1-3}$). We use this kind of cloning transformation to create genuine triangle networks\cite{renou2019limits}. In this case the cloning transformation is same as given in equation \ref{eq:B-H_gen_transform13}, with $M=64$.
\section{Three Party Quantum Networks and its Bounds and Quantum Finner Inequalities}

In this section we give a brief overview of three party networks.  We begin by considering a quantum network
with three observers $A$, $B$, and $C$. The network is created by one or more  sources by sending quantum states to the different observers. Each of these parties carries out a measurement on the
obtained quantum systems, with  measurements outputs as $a$, $b$,
and $c$. For the purpose of simplicity we are free to choose $a, b, c$ from the binary values $0$ and $1$.\\

\noindent \textbf{Three In-equivalent Networks: }\\

\begin{figure}[h]
\begin{center}
\includegraphics[scale=0.3]{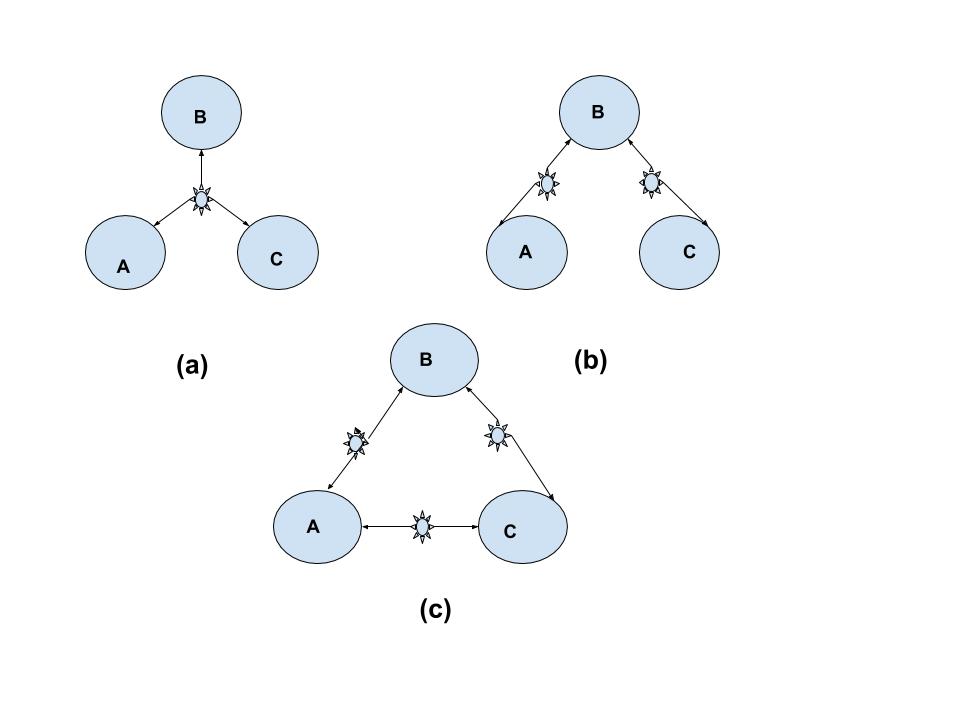}
\end{center}
\caption{\noindent \scriptsize Three Party Quantum Networks. 
 \label{fig:Network}}
\end{figure}

\begin{itemize}
\item For three observers we can have three in-equivalent networks.
The first of the kind is shown in Figure \ref{fig:Network} (a), is of a single common
source communicating a quantum state to each of these three observers. We can see that any
possible distribution $P(abc)$ can be obtained in this case. In fact, it is
enough to restrict to classical sources here. The set of possible attainable distributions $P(abc)$ is
nothing but the whole probability simplex with the normalization condition $\sum_{abc} P(abc)=1$.

\item Next we consider the case when the triangle network is created  by two independent sources, as in Figure \ref{fig:Network} (b). One of the source
distributes a common state to the parties $A$ and $B$ while the other gives it to $B$ and $C$. This is called as bi-local network.  The interesting part of this network is that observers $A$ and $C$ are initially independent, however they can be made correlated to each other through B. If we remove the node $B$ from the network, these two become completely independent. To test the independence we have the following condition\cite{renou2019limits}, 
\begin{equation}
    \sum_b P(abc) = P(a)P(c) .
    \label{eq:bilocality}
\end{equation}

%Hence, contrary to the first network discussed above, not all
%correlations are possible in the bilocality network. It turns
%out, however, that the constraint (1) is enough to characterize %achievable correlations: Any PðabcÞ satisfying (1) can
%be achieved. It is again enough to consider only classical
%variables. Specifically, let the first (respectively, second)
%source sample from PAðaÞ (respectively, PCðcÞ) and distribute the output toAandB(respectively,BandC) andBuse
%local randomness to sample PðbjacÞ. Geometrically, the set
%of achievable distributions PðabcÞ forms a six-dimensional
%curved manifold in R8.

\item Lastly, we have  the
most interesting and challenging three party network which is the triangle network
[see Figure \ref{fig:Network} (c)]. This is obtained from  bilocal network by adding a source connecting A and C. Because of this additional source, there is no need for the independence condition as it no longer holds. 
\end{itemize}

\noindent \textbf{Quantum Finner Inequality :} Quantum version of the Finner inequalities as proposed by Renou et. al. in Ref\cite{renou2019limits}, put fundamental limits on network topologies which can be generated using independent sources of entangled pairs. This inequality is mathematically expressed as,
\begin{equation}
P(abc) \leq \sqrt{P(a)P(b)P(c)},
\end{equation}
where, $P(abc)$ is the probability of observing $abc$ on combined system and $P(a)$ stands for probability of getting $a$ (same for $b$ and $c$) as measurement outcome in the subsystem $A$ similarly for subsystems $B$ and $C$ respectively.

If in a network topology we observe violation of quantum Finner inequality, we can surely say that the network is created using a single dependent source. However, in the case where there is no violation  we cannot say that these network topologies created using independent sources \cite{renou2019limits}. The network topologies we have in this article  are all generated using the process of quantum cloning. We see several instances of such networks which cannot be distinguished (that it was the network created by independent source distributing entangled pairs or was it created by cloning a single source) by the quantum Finner inequalities. 
Since quantum cloning is a unitary transformation acting on the state to be cloned along with auxiliary blank states, any state which we derive as an outcome of cloning is physical. If the state obtained exhibits entanglement, we can use them to create a quantum network. All these networks are created by the physical process of quantum cloning, that is by application of unitary transformations on initial entangled or non-entangled states, so there is no such physical restriction in the terms of achieving these topologies, which was the case when the networks were created using independent sources. However, in this case the restrictions arise in the terms of resource fullness of the networks created by cloning.\\

\noindent \textbf{Modified Finner Inequality :} We slightly modify the conditions given by the quantum Finner inequalities for bi-local networks and we observe that they become more efficient in distinguishing the cloned networks. The modification is to make observer $B$ perform two measurements, one on qubit $2$ and another on qubit $4$ instead of combined measurement on $24$. The corresponding Finner inequalities become \cite{renou2019limits}, 

\begin{equation}
    P(abb'c) \leq \sqrt{P(a)P(b)P(b')P(c)},
\label{eq:modified-finner}
\end{equation}
where $b$ is the observation outcome for qubit $2$ and $b'$ is the observation outcome of qubit $4$. The main reason why the extension performs better than the original inequality is that in the cloned bi-local network we have additional correlations like in $\rho_{14}$ and $\rho_{23}$. It is more like a four-party network but one of the observers holds two shares. 
\section{Three Party Networks By Cloning}

A typical quantum network consists of independent sources distributing entangled states to spatially separated nodes
which can then perform some operations on their part of the network. In this work, we partially relax the condition of having an independent source that distributes entangled pairs. We rather start with a single source use cloning to create similar network topologies. Several tripartite  network topologies are possible to create using quantum cloning. In the following subsections we elaborate techniques to generate various quantum networks for three parties.  
\subsection{Three copies of single qubit}
The first and the most basic method used to generate a three-party network is by creating three copies of a single qubit and distribute it to three parties. We add the restrictions that the two-qubit states, $\rho_{ab} = \rho_{bc} = \rho_{ca}$ (equality holds only in the case of symmetric cloning) must be entangled. In such a scenario, Finner inequalities do not impose any restrictions, and the entire probability simplex $P(abc)$ is possible.  
\begin{figure}[h]
\begin{center}
\includegraphics[scale=0.48]{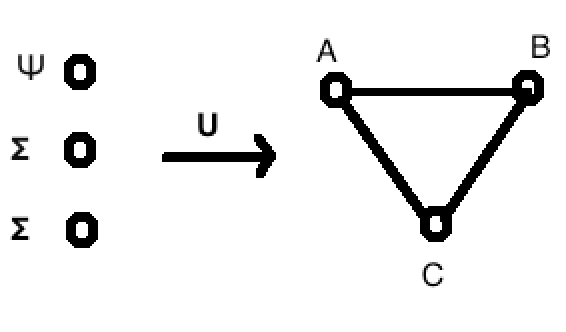}
\end{center}
\caption{\noindent \scriptsize Figure shows the cloning process to create three copies of an unknown quantum state $|\psi\rangle$ along with the result of cloning. $\sum$ represents the blank state and $U$ represents the unitary transformation for the cloning process. 
 \label{fig:Local13}}
\end{figure}
In figure \ref{fig:Local13}, we have shown the process of creating this type of network using cloning transformation given in equation \ref{eq:B-H_gen_transform13} for $M=2$. $\Psi_{ab}$ is the initial single-qubit state, $\Sigma$ is the blank state and $U$ represents the cloning unitary. This is not actually a network, rather it multipartite correlation in triangle topology.

\subsection{Local cloning of entangled pair: Bi-Local Network}
\begin{figure}[h]
\begin{center}
\includegraphics[scale=0.38]{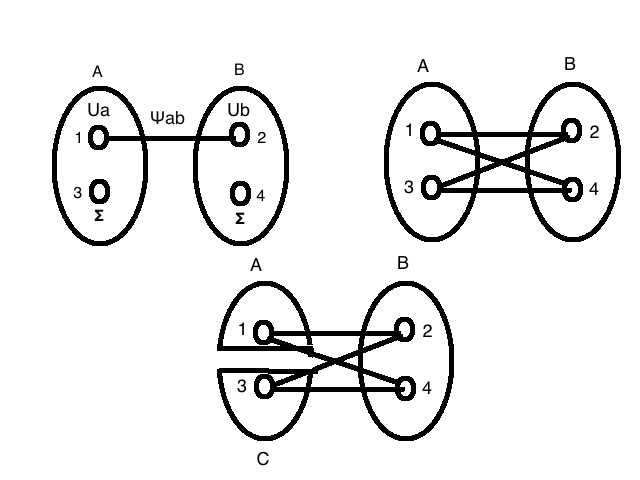}
\end{center}
\caption{\noindent \scriptsize Here we have shown the process of creating a network analogous to a bi-local network using local cloning of the initial entangled state $\psi_{ab}$. To create a three party network we send the cloned output $3$ to party $C$. $B$ keeps $2$ qubits namely $2$ and $4$. 
\label{fig:local12}}
\end{figure}

%More interesting cases arise when we want to create a bi-local network. A bi-local network is defined as a network between three parties $A$, $B$, and $C$ where $A$ and $C$ are not directly correlated but they are correlated through $B$. 
The first method we use here to create a bi-local network is by using two local cloning operations. We start initially with two spatially separated parties $A$ and $B$, which share an entangled pair $\Psi_{ab}=\alpha^2 |00\rangle +\sqrt{1-\alpha^2}|11\rangle$. Parties $A$ and $B$ apply a local cloning transformation of the form given in equation (\ref{eq:B-H_gen_transform}) on their share of the entangled pair. The transformation on the entangled pair can be given by $U_a \otimes U_b$. After cloning and tracing out the machine states, the complete output state is given by $\rho_{1234}$, where qubits $1$ and $3$ are with observer $A$ and qubits $2$ and $4$ are with observer $B$. It is important to note that the sub-states represented by qubits $\rho_{12}$, $\rho_{14}$, $\rho_{23}$, and $\rho_{34}$ are the same in case of symmetric cloning transformation which we have selected. Also, the sub-states $\rho_{13}$ and $\rho_{24}$ are the same. We have shown this process in figure \ref{fig:local12}.

We observe that under certain conditions on initial entangled state, we can simultaneously make $\rho_{12}$ (also $\rho_{14}$, $\rho_{34}$, $\rho_{23}$) entangled and $\rho_{13}$ (also $\rho{24}$) separable. To generate a bi-local network from such a configuration, $A$ can send one of the qubits, say $3$ to $C$, hence we obtain a network in which $A$ and $C$ are not directly entangled but they are entangled through $B$. It is interesting to note that whenever the qubit represented by $\rho_{13}$ is separable, the inequality given in equation \ref{eq:bilocality} is obeyed.  
 \begin{figure}[h]
\begin{center}
\includegraphics[scale=0.68]{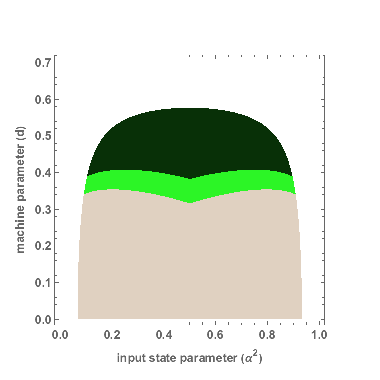}
\end{center}
\caption{\noindent \scriptsize The figure shows the states which can be distinguished in a cloned bi-local network by Finner inequality to be created by process of applying unitary on a single input source. The parameter $\alpha^2$ on the $x$ axis represents the input state parameter and $d$ on Y-axis represents the machine state parameter. The complete shaded region shows the states which are entangled in a bi-local cloned network, the darker green region represents the states which can be distinguished by original Finner inequalities and the light green region represents additional states which can be distinguished by the modified Finner inequality. The figure is generated by considering 2000 data points.
 \label{fig:finnerlocal12}}
\end{figure}
Another important aspect of the problem is distinguish-ability by Finner inequalities, we see that there are several states  which are created by local cloning and can be distinguished as being created by a single source using Finner inequalities. This is represented in figure \ref{fig:finnerlocal12}. Here, the overlapping region of all three colours (brown, light and dark green) represent those output states created by local cloning, which are entangled. Since we are using symmetric cloning this means that all the outputs namely $\rho_{12}$, $\rho_{14}$, $\rho_{23}$, $\rho_{14}$ are entangled, also, $\rho_{13}$ and $\rho_{24}$ are simultaneously separable. The dark green region in figure \ref{fig:finnerlocal12} represents those bi-local cloned networks which can be distinguished by application of Finner Inequalities. The lighter green region show additional bi-local cloned network states which can be distinguished as created by single source using modified Finner inequalities as given in equation \ref{eq:modified-finner}.

\subsection{Bi-local network: Non local cloning}
\begin{figure}[h]
\begin{center}
\includegraphics[scale=0.38]{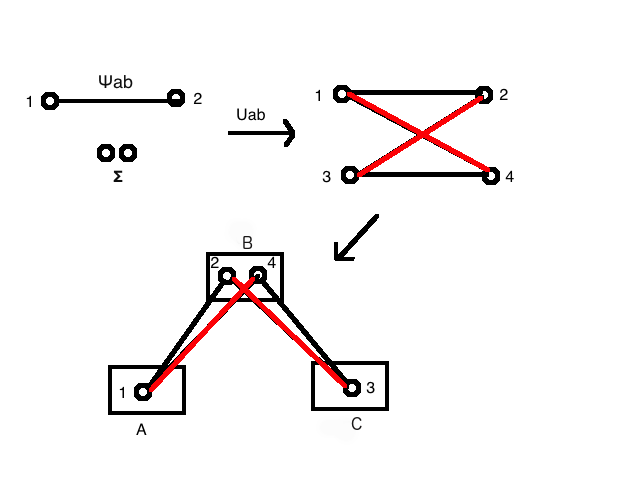}
\end{center}
\caption{\noindent \scriptsize Here we have shown the process of creating a bi-local network using non-local cloning of an initial entangled state $\psi_{ab}$. To create a three party network we send the cloned output $1$ to party $A$ and output $3$ to party $C$.
 \label{fig:nonlocal12}}
\end{figure}
Similar to the case of local cloning we can also prepare a bi-local network using non-local cloning. The difference here is that a single common source prepares the individual sources by application of non-local cloning (unitary) transformation and then distributes qubits to the three parties involved to create a network. In this case, an independent third party has an entangled pair of particles, it applies a global unitary transformation $U_{ab}$ and creates a copy of the entangled pair. In this case as well we have $\rho_{12}$, $\rho_{14}$, $\rho_{23}$ and $\rho_{34}$ are equal and $\rho_{13}$ is equal to $\rho_{24}$. It is important to note here that the correlations are better copied in the case of non-local cloning, so we might actually be able to create a bi-local network with a wider range of input states. 

The source after performing the cloning transformation sends the qubit $1$ to observer $A$, qubit $3$ to $C$, and qubits $2$ and $4$ to $B$. It is important to note that under certain selected input state parameters and machine state parameters we can obtain the condition where qubits $1$ and $3$ are separable, also qubits $2$ and $4$ are separable.
In some overlapping range of input state parameters and machine state, we can also have entanglement between $1,2$ and $1,4$ and also among $2,3$ and $3,4$. Hence we establish the condition of bi-locality. 

\begin{figure}[h]
\begin{center}
\includegraphics[scale=0.68]{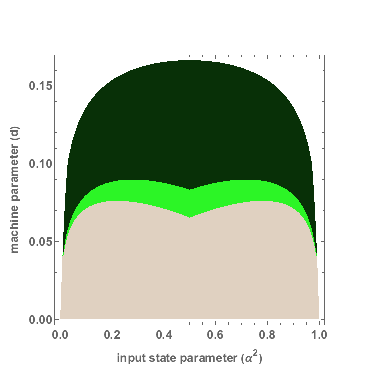}
\end{center}
\caption{\noindent \scriptsize The figure shows the states which can be distinguished as being created by a single source in a cloned bi-local network by Finner inequality. The parameter $\alpha^2$ on the $x$ axis represents the input state parameter of the state to be cloned and $d$ on Y-axis represents the cloning machine state parameter. The complete brown region shows the states which are entangled in a bi-local cloned network, the dark green region represents the states which can be distinguished by original Finner inequalities, and the larger dark and light green regions combined represent the states which can be distinguished by modified Finner inequality. The figure is generated by considering 2000 data points.
 \label{fig:finnerNonLocal12}}
\end{figure}

Focusing on the distinguish-ability of these networks using Finner inequalities. We see that there are networks which cannot be distinguished as created by single-source using cloning or an independent source distributing entangled pairs. These networks are represented in figure \ref{fig:finnerNonLocal12}.The complete shaded region shows all states where the cloning outputs are entangles and can be used for creating a network, the dark green region represents those bi-local cloned networks which can be distinguished by application of Finner Inequalities. The lighter green region show additional bi-local cloned network states which can be distinguished as created by single source using modified Finner inequalities as given in equation \ref{eq:modified-finner}.   

\subsection{Triangle network: Non local cloning}
The last network topology we consider is the extended triangle network. The extension is in terms of additional internal correlations which exist among the qubits held by the observers $A$, $B$ and $C$. Here, we start with an independent entangled pair with a third party, and apply a cloning transformation of type given in eq (\ref{eq:B-H_gen_transform13}) with $M=4$, to obtain three copies of the entangled pairs. In this case other than the intended entanglement between $\rho_{12}$, $\rho_{34}$ and $\rho_{56}$, we also have  $\rho_{14}$ = $\rho_{23}$ = $\rho_{16}$ = $\rho_{25}$ = $\rho_{36}$ = $\rho_{45}$ which are entangled for some specific machine state parameter and input states. On the other hand,
$\rho_{13}$ = $\rho_{15}$ = $\rho_{35}$ =  $\rho_{24}$ = $\rho_{26}$ =  $\rho_{46}$ remain largely separable. 

\begin{figure}[h]
\begin{center}
\includegraphics[scale=0.38]{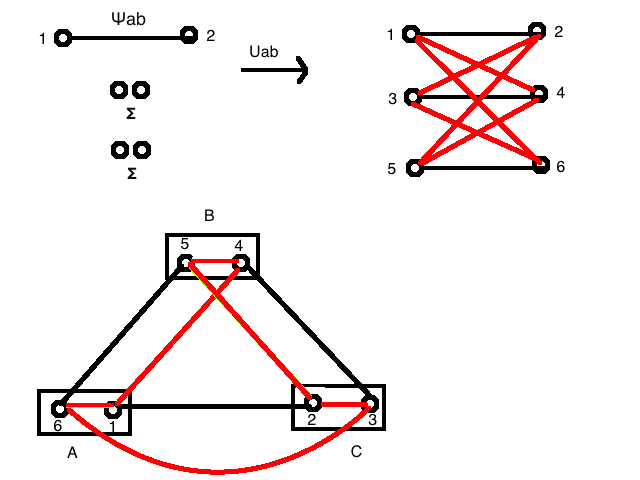}
\end{center}
\caption{\noindent \scriptsize Here we have shown the process of creation of a network analogous to the triangle network using non-local cloning, creating two additional copies of an initial entangled state $\psi_{ab}$. To create a three-party network we send the cloned output $1, 6$ to party $A$, and output $2,3$ to party $C$. The Red color lines represent the additional correlations present in cloned networks that are not in originally described triangle networks.
 \label{fig:nonlocal13}}
\end{figure}

The network which is created by following this strategy is not a genuine triangle network, because it has other entangled states as well. The method of preparing such a network is shown in figure \ref{fig:nonlocal13}. The solid lines represent entangled states, the red color, and the black color are used to demarcate two different types of states. It would be interesting to see, if we can obtain the genuine triangle networks by finding a condition on machine state parameter and the input states which can make $\rho_{14}$ (also $\rho_{23}$, $\rho_{16}$, $\rho_{25}$, $\rho_{36}$, $\rho_{45}$) separable and keep $\rho_{12}$ ($\rho_{34}$ and $\rho_{56}$ entangled. 

\begin{figure}[h]
\begin{center}
\includegraphics[scale=0.68]{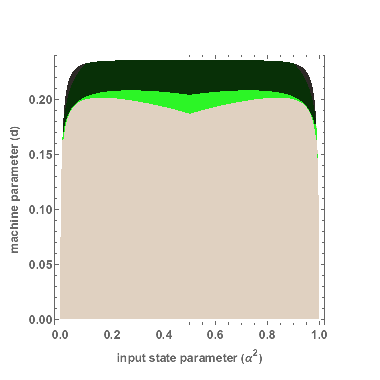}
\end{center}
\caption{\noindent \scriptsize The figure shows the states which can be distinguished in a cloned triangle network by Finner inequality to be created by a single source. The parameter $\alpha^2$ on the $x$ axis represents the input state parameter and $d$ on Y-axis represents the machine state parameter. The complete shaded region shows the states which are entangled in a bi-local cloned network, the darker green region represents the states which can be distinguished by original Finner inequalities and the light green region represents the states which can be distinguished by modified Finner inequality. The figure is generated by considering 2000 data points.
 \label{fig:finner16}}
\end{figure}

Focusing on the distinguish-ability of these networks using Finner inequalities. We see that there are networks that cannot be distinguished as created by single-source using cloning from independent sources distributing entangled pairs. These networks are represented in figure \ref{fig:finner16}.  The parameter $\alpha^2$ on the $x$ axis represents the input state parameter and $d$ on the Y-axis represents the machine state parameter. The complete shaded region shows the states which are entangled in a triangle cloned network, the darker green region represents the states which can be distinguished by original Finner inequalities and the light green region represents the additional states which can be distinguished by modified Finner inequality.

\section{Witnessing Entanglement in Triangle Networks}
In this section, we investigate quantum
correlations in networks from the point of view of tripartite mutual information. We also use the notion of network-based entanglement witness as defined in \cite{kraft2020quantum} to check if the networks created by cloning belong to $\Delta_i$, where $\Delta_i$ represents the quantum networks created by distributing entangled pairs created by independent sources. We use this witness to quantify the amount of violation and hence quantify the amount of dependence a source has in case of networks generated with cloning. For this study, we restrict ourselves to only genuine triangle network topology, the extension to other typologies which we have talked above is straightforward. 

In the study by Kraft et. al \cite{kraft2020quantum}, several observations are shared which can be used to detect if a state $\rho$ belongs to $\Delta_i$, namely,
\begin{itemize}
   \item 1. $I_3(A : B : C)$ = $0$ for any $\rho \in \Delta_i$. Here, $I$ represents tripartite mutual information give by,
  \begin{eqnarray}
&& I_3(A : B : C) = S(ABC) + S(A) + S(B) + S(C) \nonumber\\
&& - S(AB) - S(AC) - S(BC).
\label{eq:TMI}
\end{eqnarray}
  Intuitively this means that there is no shared classical mutual information within the subsets of the quantum states owned by the parties $A$, $B$ and $C$ respectively.
  
  \item 2. Let $E[\rho]$ be an entanglement measure
that is additive on tensor products and monogamous.
For any $\rho \in \Delta_i$ we have that $E_{X|Y Z}[\rho] = E_{X|Y} [tr_Z \rho] +
E_{X|Z}[tr_Y \rho]$ holds for all the bipartitions $A|BC$, $B|AC$, and $C|AB$. The intuition here is that the entanglement
on the bi-partition $A|BC$ should be equal to the sum of
the entanglement in the reduced states, i.e. $A|B$ and
$A|C$.
\end{itemize}
We have used squashed entanglement $E$ \cite{christandl2004squashed} as a measure to quantify entanglement. Squashed entanglement guarantees additivity and monogamy properties. It is defined as

\begin{equation}
    E_{sq}(\rho^{AB}) := inf \{ \frac{1}{2}I(A;B|E):\rho^{ABE} extension of \rho^{AB} \}.
\end{equation}
Here, minimum is taken over all extensions of $\rho^{AB}$ given by $\rho^{ABE}$, such that $\rho^{AB} = Tr_E[\rho^{ABE}]$. $I(A;B|E)=$  
$S(AE)$ + $S(BE)$ - $S(ABE)$ - $S(E)$ is the quantum conditional mutual information of $\rho^{ABE}$\cite{cerf1997negative}.

We have considered cloned triangle networks for the analysis of  observations $1$ and $2$. We first consider the observation $1$, and test if the cloned triangle network $\rho^{ABC}$ fall into $\Delta_i$, i.e. set of triangle networks created by independent sources and then we use observation $2$ for the cases where  $\rho^{ABC} \notin \Delta_i$ to quantify the amount of dependence which each entangle pair have in such a network. 

\begin{figure}[h]
\begin{center}
\includegraphics[scale=0.68]{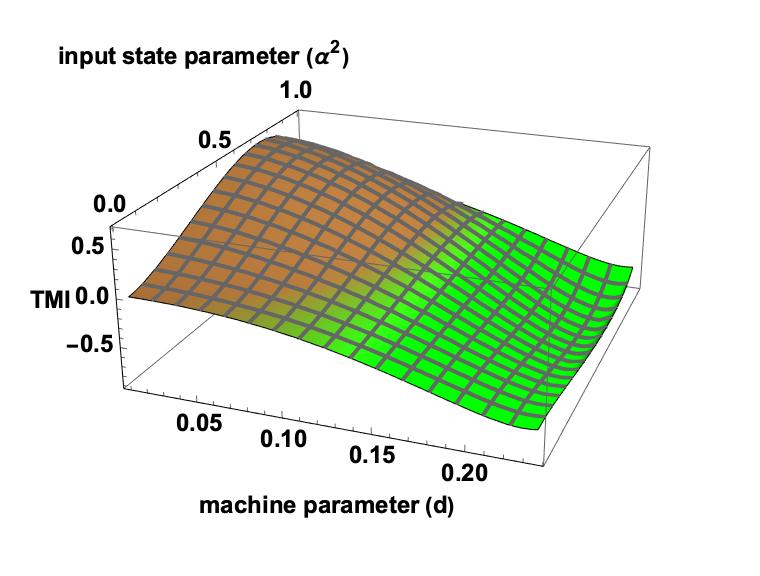}
\end{center}
\caption{\noindent \scriptsize Here we have shown the values of tripartite mutual information (TMI) of various cloned triangle networks given by input state parameter $\alpha$ and machine state parameter $d$. We note that both positive and negative values of TMI are obtained. The figure is generated by considering 2000 data points.
 \label{fig:TMI}}
\end{figure}

In figure \ref{fig:TMI} we have shown the values for tripartite mutual information for various states $\rho_{ABC}$ obtained after cloning (creating two copies) non-maximally entangled state, $\ket{\psi} = \alpha\ket{00} + \sqrt{1-\alpha^2}\ket{11}$. Here, $d$ represents the machine state parameter for the cloning machine. We see that both positive and negative values for the $TMI$ are obtained. If $TMI < 0$, mutual information is monogamous like entanglement measures. We see that there are no such states where $TMI$ is exactly zero, hence all cloned triangle networks can be verified of not being part of $\Delta_i$.

We define amount of dependence in cloned triangle networks as,
\begin{equation}
   D[\rho^{ABC}] = |E_{A|BC}[\rho^{ABC}] - E_{A|B} [tr_C \rho^{ABC}] -
E_{A|C}[tr_B \rho^{ABC}]|.
\end{equation}

\begin{figure}[H]
\begin{center}
\includegraphics[scale=0.43]{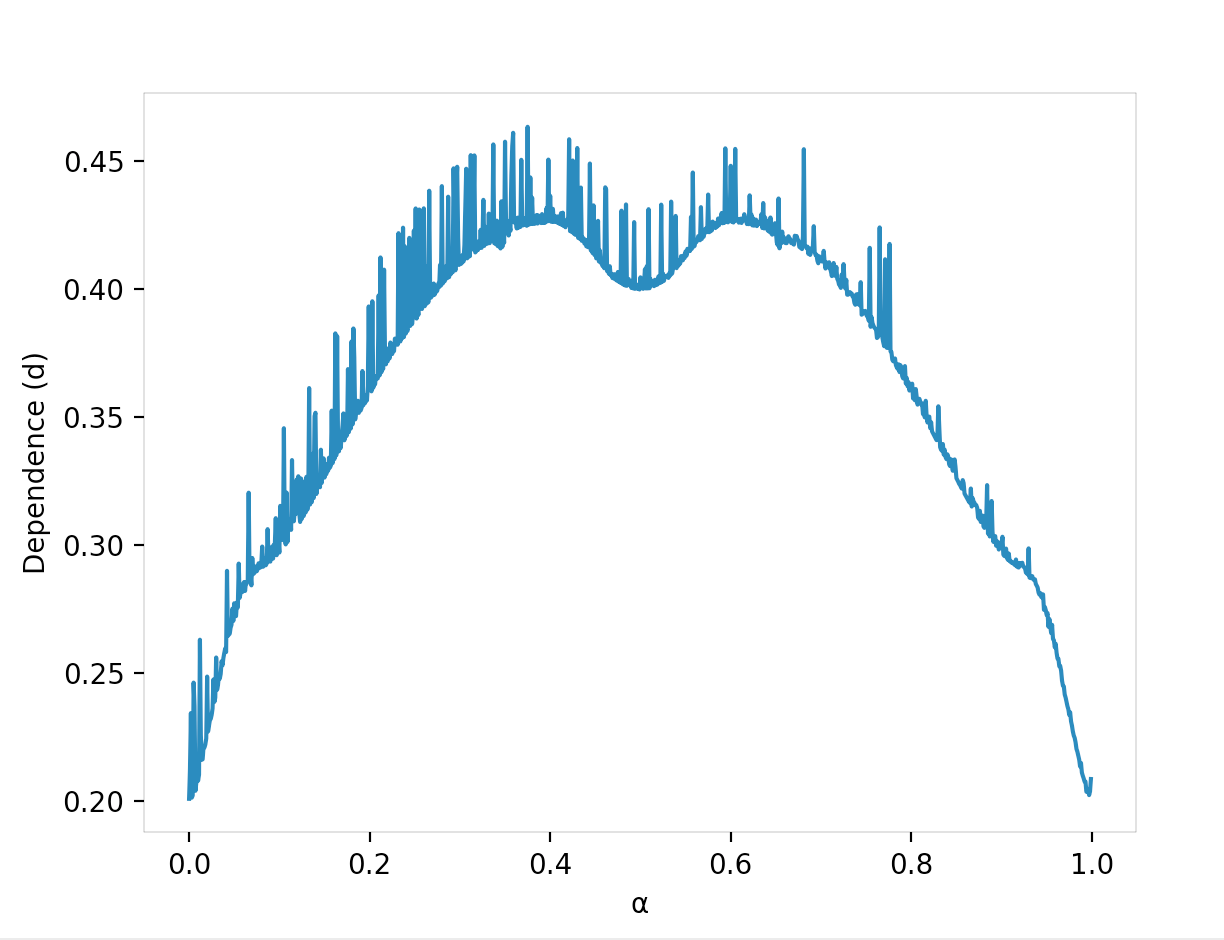}
\end{center}
\caption{\noindent \scriptsize The figure shows the amount of statistical dependence on the initial state which prepares the network through cloning. The parameter on x axis ($\alpha$) represents the input state parameter for non-maximally entangles states and on y-axis, the dependence values are marked. Here the machine cloning machine parameter $d$ is set equal to $0.1$.
 \label{fig:SqENT}}
\end{figure}

Using this, we now have a quantitative measure of how much dependency on the initial variable $\alpha^2$, which was used by the source to create the non-maximally state is left in the cloned triangle networks. In figure \ref{fig:SqENT} we have shown the dependence values present in non maximally entangled states as a function of state parameter $\alpha$. Here for simplicity, the machine state parameter is set to 0.1. To generate this we have calculated squashed entanglement in state $\rho_{123456}$ which represents the cloned triangle network given in figure \ref{fig:nonlocal13}, this reduces to the calculation of

\begin{equation}
   D[\rho^{123456}] = |E_{A|BC}[\rho^{123456}] - E_{A|B} [ \rho^{1456}] -
E_{A|C}[\rho^{1236}]|.
\end{equation}
 
\section{Conclusion}
In this work, we have created various tripartite  network topologies with the help of cloning a bipartite two-qubit state. We have used several cloning techniques like local and non-local with two copies and three copies to create various types of cloned networks. Creation of quantum networks is of importance because in all those situations where the entanglement cannot be generated we can use the cloning process to obtain a larger network of correlated qubits. We explore the Finner inequalities to check if the obtained cloned networks can be distinguished from networks created by independent sources distributing entangled pairs. We note that some cloned networks cannot be distinguished. We further explore the tripartite mutual information and squashed entanglement to quantify this dependence for triangle networks. \\

\noindent \textit{Acknowledgement:} This work is supported by 
National Natural Science Foundation of China under Grant No.
61877054, 12031004, 11901526 and 11571307. I. Chakrabarty and M.
Shukla thank Zhejiang University for supporting their visits.


%\begin{widetext}

\end{document}